# Can Air Pollution Save Lives? Air Quality and Risky Behaviors on Roads

Wen Hsu, Bing-Fang Hwang, Chau-Ren Jung, Yau-Huo (Jimmy) Shr[†]

October 9th, 2021

## Abstract

Air pollution has been linked to elevated levels of risk aversion. This paper provides the first evidence showing that such effect reduces life-threatening risky behaviors. We study the impact of air pollution on traffic accidents caused by risky driving behaviors, using the universe of accident records and high-resolution air quality data of Taiwan from 2009 to 2015. We find that air pollution significantly decreases accidents caused by driver violations, and that this effect is nonlinear. In addition, our results suggest that air pollution primarily reduces road users' risky behaviors through visual channels rather than through the respiratory system.

Keywords: air quality, risk preferences, traffic accidents

JEL Code: Q53, I12, R41

[†] Hsu Wen is a research associate in the Department of Agricultural Economics at National Taiwan University. Bing-Fang Hwang is a professor of Occupational Safety and Health at China Medical University. Chau-Ren Jung is an assistant professor of Public Health at China Medical University. Yau-Huo (Jimmy) Shr is an assistant professor of Agricultural Economics at National Taiwan University (yhshr@ntu.edu.tw). The individual traffic accident data used in this study are from the National Police Agency of Taiwan. The data can be obtained by filing a request directly with the National Police Agency. Authors are willing to assist those interested in filing the request. The aggregated data used for all analyses in the study will made available in the Online Appendix upon acceptance. Huang and Shr gratefully acknowledge funding from the Ministry of Science and Technology (MOST 108-2621-M-039-001 & 109-2410-H-002-157). The authors have no relevant financial or material interests related to the research described in the paper.

# I. Introduction

Numerous studies have documented the effects of air pollution on health and various societal outcomes, and most, if not all, of those effects are adverse (for a recent review, see Currie and Walker 2019). In this study, we ask whether air pollution can reduce life-threatening risky behaviors through greater risk aversion.

The inquiry arises from the evidence showing the adverse effects of air pollution on cognitive performance and labor productivity (Archsmith et al. 2018, Bedi et al. 2021, Carneiro et al. 2021, Chang et al. 2019, Dong et al. 2021, Graff Zivin and Neidell 2012, He et al. 2019, Heissel et al. 2020, Lai et al. 2021, Lichter et al. 2017, Persico and Venator 2021), and that cognitive ability is negatively associated with the degree of risk aversion (Dohmen et al. 2010, Dohmen et al. 2018).[1] Indeed, a series of papers, focusing on financial investment behaviors, have found that elevated air pollution leads to a higher degree of risk aversion and lower returns (Heyes et al. 2016, Levy and Yagil 2011, Li et al. 2019). In an economic laboratory experiment setting, Chew et al. (2021) find that higher concentrations of $PM_{2.5}$ (fine particulate matter) increases risk and ambiguity aversions. Medical literature has also provided corroborating evidence showing the positive relationship between air pollution levels and the degree of risk aversion (Li et al. 2017).

In this study, we investigate the contemporaneous effect of air pollution on traffic accidents with casualties caused by any drivers involved in committing a traffic violation, a risky behavior that poses immediate threat to lives. Our analyses leverage the administrative traffic accident records of Taiwan for the years 2009 through 2015, which includes more than 1.1 million such accidents, as well as high-resolution air quality data integrating satellite, ground monitoring, long-term emission, and land use data. To overcome the potential endogeneity between traffic accidents and air pollution, we use wind directions to induce exogenous variations in air quality. The results confirm our hypothesis. Based on our preferred specification, we find that a 1 $\mu g/m^3$ increase in $PM_{2.5}$ concentration decreases the number of accidents caused by driver violations by 0.59%. This negative effect is robust to a host of different specifications.[2]

---

[1] Schmidt (2019) reviews studies on the effect of air pollution on productivity. For a review of how air pollutants affect brain functions, see Tsai et al. (2019).

Air pollution could also lead to more traffic accidents through impaired cognition or irritated respiratory organs, two mechanisms which have been highlighted in the literature on the adverse effects of air pollution on cognitive performance and productivity (Sager 2019).[3] Therefore, the existing evidence suggests that air pollution can affect road safety in two opposite ways: by decreasing traffic accidents through increased degree of risk aversion, and by increasing accidents through impaired cognition. This is particularly relevant when the dose-response relationships between air pollution and risk attitude/cognitive performance are nonlinear so that the combined effect on road accidents can be either positive or negative. To shed light on the effect of air pollution on risky behaviors, we focus on accidents caused by traffic violations.[4]

This study further explores the transmission channels through which air quality influences risk preferences and associated behaviors. Inhalation exposure has been recognized as the primary route of air pollution affecting humans (e.g., Brook et al. 2002 and Gurgueira et al. 2002). Li et al. (2017) find that inhalation of air pollutants leads to an increase in the stress hormone cortisol, which has been linked to higher levels of risk aversion (Coates et al. 2008, Kandasamy et al. 2014, Rosenblitt et al. 2001). It is plausible to postulate that inhalation again is the main channel for air quality to change people's risk preferences and risky behaviors. Still, Lu et al. (2018) find that U.S. students show higher level of anxiety by simply looking at pictures of polluted air. Higher levels of (fine) particulate matter, $NO_x$, and $SO_2$ can lead to higher light attenuation and soften the sky color, so the sky looks gray or hazy (Huang et al. 2014, Liu et al. 2020, Ouimette and Flagan 1982). Therefore, although not scientifically precise, levels of air pollution can be visually assessed. More importantly, it is highly likely that the majority of citizens judge air quality based on what they "see" in the air (Middleton et al. 1985, Van Brussel and Huyse 2019).[5] Therefore, we hypothesize that elevated air pollution levels can also affect emotions, risk preferences, and risky behaviors through visual channel. Indeed, this potential visual channel is consistent with the literature in both economics and psychology showing that gloomy or smoky sky makes humans more risk averse or pessimistic (Bassi et al. 2013, Goetzmann et al. 2015, Hirshleifer and Shumway 2003, Kliger and Levi 2003, Saunders 1993).

---

[2] Many studies have used wind directions to instrument air quality (e.g., Anderson 2020, Bondy et al. 2020, and Deryugina et al. 2019). Note that, instrumental variable addresses biases stemming from avoidance behavior against air pollution, in addition to other issues such as measurement errors (Moretti and Neidell 2011). Therefore, the negative effect is unlikely explained by avoidance behaviors such as staying at home when air pollution levels are high. As discussed later, we perform auxiliary analyses and find no evidence showing that avoidance behavior could be a credible competing explanation for the reduced number of accidents.

[3] Sager (2019), the first economic study focusing on the effect of air quality on road safety to our knowledge, finds elevated $PM_{2.5}$ concentration increases the number of vehicles involved in *all* accidents with casualties in the United Kingdom. The effect of cognitive impairment on driving performance has already been widely documented in different disciplines such as public health and neurological science (e.g., Anstey et al. 2012 and Schultheis et al. 2001).

[4] As a placebo test, we also examine the effect of air pollution on accidents caused by errors (harmless lapses), on which we hypothesize that risk aversion should have smaller impact.

With fewer physical barriers to air pollutants on roads, drivers of non-enclosed vehicles (e.g., motorcyclists, scooter riders, and bikers) have been found to be exposed to higher levels of air pollution, compared to drivers of enclosed vehicles (e.g., car and truck drivers) in Taiwan and many other countries (Dirks et al. 2012, Rivas et al. 2017, Tsai et al. 2008, Vlachokostas et al. 2012). In an experiment conducted by Taiwan Environmental Protection Agency, the exposure to $PM_{2.5}$ among motorcycle/scooter riders is found to be four times higher than that among car drivers in a given period of time (Taiwan EPA, 2017). If air pollution can affect risky behaviors on roads through the respiratory system, we would expect to see the effect to be stronger on non-enclosed vehicle drivers than on enclosed vehicle drivers.[6] Therefore, to pin down whether such respiratory channel exists, we test the heterogeneous effects of air pollution on accidents with an enclosed/ non-enclosed vehicle driver being principally liable.

---

[5] In a survey on citizens' attitudes and perceptions of air quality in Taiwan, more than 70% of the respondents stated that they primarily assess contemporaneous air quality by how hazy the sky is (Shr et al. 2021).

Air pollutants can be visible because their effect on light absorbance and the Molar light attenuation coefficient (Brimblecombe 1987). When there is little ambient light, air pollution will have little effect on light attenuation and be much less visible (Hyslop 2009, Yu et al. 2018). Therefore, road users arguably can only visually assess air quality in time with natural light, assuming air pollution is not bad enough to create severe ground-level smog and low-visibility conditions for driving.[7] Therefore, to examine if air quality affects risky behaviors on road through visual channel, we investigate the heterogeneous effect on accidents caused by violations during time with and without natural light.

Surprisingly, we find no discernible differences between the effects on the number of accidents caused by violations committed by non-enclosed/enclosed vehicle drivers. However, the negative effect of air pollution on accidents is only observed during times with natural light. These results show that air pollution predominately affects risky road behaviors through visual channels, and there is no evidence showing that the respiratory system plays a role.

---

[6] With one of the highest numbers of motorcycles and scooters per adult (0.71/person) and densities (391/km$^2$) in the world, non-enclosed vehicle drivers are representative of the entire adult population in Taiwan. We argue that any heterogeneous effects of air pollution across enclosed and non-enclosed vehicle drivers are less likely to be attributed to cross-group heterogeneity in socio-demographics.

[7] Maurer et al. (2019) find that, when excluding foggy conditions, visibility was never lower than 2 km in Taiwan between 1986 and 2016, regardless of the $PM_{10}$ concentration. In the analysis, we examine if our results are robust to excluding accidents under foggy (smoggy) conditions.

Overall, this paper contributes to the literature on at least four important fronts. First, we provide new evidence showing that air pollution can discourage risky behaviors in addition to those associated with financial investment. This also contributes to the line of studies on the determinants of risky behaviors (e.g., Dohmen et al. 2011 and Schoemaker 1993). Second, our results highlight the role of risk attitudes when estimating the costs of air pollution stemming from impaired cognitive performance. If the effects of air quality on risk attitudes and cognition affect the outcomes of interest in opposite directions, the adverse effect on cognitive function may be underestimated, when the effect on risk aversion is not controlled for. Our findings also provide new empirical evidence suggesting the transient malleability of risk preferences due to exogenous shocks (Schildberg-Hörisch 2018) as well as that cognitive ability and risk aversion interrelatedly affect economic outcomes at individual level (Dohmen et al. 2010, Heckman et al. 2006). Third, using data from Taiwan, our results enhance the external validity of the studies quantifying the effect of air quality on road safety. The external validity is particularly relevant given that we find such effect is nonlinear and more than 90% of the world's traffic fatalities occur in low- and middle-income countries where the air pollution levels are high (World Health Organization 2018). Last but not least, we provide, to our knowledge, the first evidence to demonstrate the transmission channels through which air quality affects risky behaviors. As we discuss later, this finding has important implications for studies quantifying the economic costs from different air pollutants.

The remainder of this paper is organized as follows. Section II presents the empirical strategy, particularly the instrumental variable approach for identifying the casual impacts of air pollution on traffic accidents caused by violations. In Section III, we describe the administrative traffic accident data and how the grid air quality data is constructed. Section IV presents the main results on the effect of air pollution on risky road behaviors, followed by a host of robustness checks. In Section V, we present the evidence for showing the transmission channel through which air pollution affects risky behaviors on roads as well as the nonlinear effect of air pollution on behaviors. Section VI concludes and discusses implications for future research.

## II. Empirical Strategy

When identifying the casual relationship between air quality and risky driving behaviors, one key source of the potential endogeneity is the failure to account for common factors such as traffic volume, weather conditions, and regional characteristics, which can affect air pollution, driving behaviors, and traffic accidents simultaneously. For example, higher traffic volumes may concurrently lead to higher levels of air pollution and increase the number of accidents (Dastoorpoor et al. 2016, Zhou and Sisiopiku 1997). In this case, omitting traffic volume will overestimate the effect of air pollution on accidents caused by risky behaviors.[8] Similarly, failure to control for weather conditions could also lead to biased effects of air pollution, because many weather variables, such as rain, humidity, temperature, and wind speed, have been found to be associated with air pollution, traffic volume, driving behaviors, and the risk of traffic accidents (Brijs et al. 2008, Caliendo et al. 2007, Wan et al. 2020).

Our first empirical strategy relies on high-dimensional fixed effects and atmospheric controls to remove confounding effects from regional- and time-invariant factors as well as weather conditions. To accommodate the large zero counts of daily accidents in each region, we estimate Poisson models with pseudo-maximum likelihood (PPML) of the following form (Silva and Tenreyro 2006):

---

[8] Traffic monitoring data at the individual road and intersection level is only available for a few major cities in Taiwan. More importantly, aggregating such fine spatial level observations into regional levels is in general not advised (Tao et al. 2004). We therefore do not attempt to explicitly to control for traffic volume and instead rely on econometric approaches to mitigate this omitted variable issue.

$$Acc_{it} = \exp(\beta AQ_{it} + \boldsymbol{\theta}'\boldsymbol{X}_{it} + \boldsymbol{\omega_{it}}) + \varepsilon_{it} \tag{1}$$

Here, $Acc_{it}$ is the number of traffic accidents caused by driver violations in region $i$ at time $t$ (which corresponds to a day, in our case); $AQ_{it}$ is the air quality measure; $\boldsymbol{X}_{it}$ is a vector of atmospheric condition controls, including temperature, precipitation, number of rain hours, relative humidity, and windspeed; $\boldsymbol{\omega_{it}}$ is a vector of spatial and temporal (year-by-month and day-of-week) fixed effects. The $\beta$ is therefore the marginal effect of a unit change in air quality on the percentage change in the associated number of traffic accidents.[9]

In addition to models including the spatial and temporal fixed effects separately, we also estimate models with region-by-year-by-month and region-by-day-of-week fixed effects to control for factors such as traffic volume, with seasonal and within-week variations for each region. For example, a region with more businesses may have more traffic during weekdays than on weekends, but the within-week distribution of traffic volume may be different in regions with more recreation resources. Including the interactions between spatial and temporal fixed effects can control for such dynamics by allowing the systematic time-varying effects to be region-specific. Note that, by including regional-by-year-by-month fixed effects, the models have also controlled for some common region-specific time-varying factors such as population. Standard errors are clustered at both day and district/township levels to account for potential temporal and spatial autocorrelations.

Ignoring avoidance behaviors such as when some individuals decide not to travel because of a high level of air pollution would bias the effects of air pollution on health outcomes downward (Neidell 2009). Moretti and Neidell (2011) have employed an instrumental variable approach to account for avoidance behavior, omitted variable, and measurement error in pollution exposure, when estimating the effect of ozone on hospitalization. Reverse causality is another concern. Although our main hypothesis is that air pollution affects the number of accidents caused by risky driving behaviors through altered risk attitudes, the occurrence of accidents may create congestion and increase air pollution or extend the time that a driver is exposed to air pollution. Studies have identified the association between traffic congestion and accidents (Green et al. 2016, Knittel et al. 2016, Pasidis 2017, Quddus et al. 2009). Therefore, simply regressing the number of traffic accidents on the level of air pollution using Ordinary Least Squares (OLS) may still result in biased estimates even without the concern of omitted variable bias.

---

[9] The PPML models are estimated using Stata package `ppmlhdfe` (Correia et al. 2020).

To address the above issues, we use an instrumental variable (IV) approach to alleviate the potential endogeneity. Wind direction has been documented as a key determinant of air quality in Taiwan: air pollutants are more likely to be deposited in downwind areas, and leeward-side areas would have higher level of pollution because of lower atmospheric dispersion (Shie et al. 2016). In addition, to our knowledge, there is no evidence showing that wind direction can directly influence risk attitudes or the number of traffic accidents. That is, the exclusion restriction is likely to hold when weather covariates and fixed effects are controlled for. A number of economic studies have also used wind directions as IVs to induce exogenous variations in air quality, when identifying its effect on various societal outcomes. For example, Anderson (2020) and Deryugina et al. (2019) focus on health outcomes and medical expenditure, He et al. (2019) examine work productivity, Li et al. (2019) look at investment behavior, and Bondy et al. (2020) focus on crime. Therefore, we use wind directions as the IVs for air quality.[10]

---

[10] Thermal (temperature) inversion is another candidate for instrumenting air quality; for example, Arceo et al. (2016) on infant mortality and Sager (2019) on road safety. Bondy et al. (2020) also use thermal inversion as another IV to corroborate the results based on wind directions. However, thermal inversion rarely occurs in summer days during

We estimate the following first-stage model:

$$AQ_{it} = \gamma' \mathbf{1}[AQZ_i = z]\mathbf{WD}_{it} + \delta' X_{it} + \varphi_{it} + v_{it} \qquad (2)$$

In the model above, $\mathbf{WD}_{it}$ is a vector of numbers of hours in a day that the wind blows from the south, west, and north in region $i$.[11] The variable $\mathbf{1}[AQZ_i = z]$ is a dummy variable indicating if region $i$ is in air quality zone $z$.[12] $\gamma$ is therefore a vector of air-quality-zone-specific marginal effects of one additional hour of a certain wind direction on air quality. By allowing the marginal effects to vary across the four air quality zones, we mitigate the possibility of violating the monotonicity assumption.

To accommodate the (nonlinear) PPML models in the second stage, we apply the control function approach by inserting the residuals, $\widehat{v_{it}}$, from the first stage (Blundell and Powell 2003, Henderson and Souto 2018, Wooldridge 2015). The corresponding equation is:

$$Acc_{it} = \exp(\beta AQ_{it} + \widehat{v_{it}} + \theta' X_{it} + \omega_{it}) + \varepsilon_{it} \qquad (3)$$

---

our study period (2009 – 2015). Also, because of Taiwan's complex terrain, the occurrence of thermal inversion varies widely across regions (Liou et al. 2004). For example, in northern Taiwan, thermal inversion occurs in less than 6% of days in our study period. Therefore, we consider thermal inversion is not an ideal candidate of IV for air quality in Taiwan.

[11] East wind serves as the reference wind direction. In addition, we test the specifications with eight wind directions (i.e., north, northeast, east, southeast, south, southwest, west, and northwest), and the results are robust.

[12] Taiwan EPA classifies the counties in Taiwan into several air quality zones, based on geographical adjacency and similarity in the relationships between air quality and atmospheric conditions.

Our key parameter of interest is therefore $\beta$ in equation (3) above. $X_{it}$ and $\omega_{it}$ are again vectors of atmospheric condition controls as well as spatial and temporal fixed effects as described in equation (1).

## III. Data

This paper uses administrative records on individual traffic accidents involving casualties between 2009 and 2015, provided by the National Police Agency under a data-sharing agreement. Each record includes key accident information, including, but not limited to, the primary contributory factor, exact timing, location (address and geo-coordinates), weather, light condition, junction detail, road type, road surface condition, and presence of road hazards, as well as information on drivers and passengers such as age, gender, occupation, type of vehicle, and whether a driver was under the influence. We aggregate the individual records to daily counts of accidents caused by violations, according to the primary contributory factor, within each of the 365 districts and townships from the 19 cities and counties in the main island of Taiwan. Panel A of table 1 presents the selected summary statistics of daily accident counts in each district and township. Among all accidents caused by driver violations, non-enclosed vehicle drivers were the most liable for 47.45% of the accidents,[13] 69.95% occurred during times with ambient natural light, 82.24% happened during fair-weather days (in contrast to cloudy, rainy, foggy, hazy, and strong wind days), 42.88% in rush hours (7AM – 10AM and 5PM – 8PM), 66.11% at intersections, and 32.49% with female drivers being the most liable party in the accidents.

The daily average air quality, measured by PM$_{2.5}$ concentration, of each district and township is calculated based on the spatial average from the 3km*3km grid-level data integrating hourly ground level monitoring and long-term ground emission data from Taiwan EPA,[14] satellite data of aerosol optical depth (AOD) retrievals from MODerate resolution Imaging Spectroradiometer (MODIS), and land use data from the National Land Surveying and Mapping Center.

---

[13] Note that this highlights the representativeness of non-enclosed vehicle drivers of all drivers on the road in Taiwan.

An empirical challenge for many economic studies focusing on the (contemporaneous) effects of air pollution is to obtain the air quality data at the same spatial resolution with outcome or other key control variables. Many recent studies rely solely on ground monitoring data and use spatial interpolation techniques, most commonly inverse distance weighting (e.g., Bondy et al. 2020, Deryugina et al. 2019) and kriging (e.g., Burkhardt et al. 2019), to tackle the spatial sparsity of ground monitoring stations. However, such spatial interpolation methods virtually ignore the roles of terrain, land use, and ground emission. This problem is particularly relevant in our study because of the highly heterogeneous landscapes and mountainous terrain in Taiwan (Jung et al. 2018). To overcome this problem, we first build a linear mixed effect model that combines the 3km resolution satellite-based AOD, land use, meteorological data to estimate $PM_{2.5}$ concentrations with comprehensive spatial coverage of Taiwan at the 3km*3km level.[15] The estimates are subsequently validated by the ground monitoring data using a ten-fold cross validation and multiple imputation with additive regression models to address the measurement error and temporal missingness of satellite data.[16] We therefore argue that our air quality data is also of better quality than data that solely relies on satellite images. For more detail regarding the methods

---

[14] The ground monitoring data are publicly available at the open data port of Taiwan EPA (https://data.epa.gov.tw/). We retrieve the ground emission data from the Taiwan Emission Data System (TEDS 8.0 and 9.0 at https://teds.epa.gov.tw/TEDS.aspx).

[15] Some key land use categories are residential areas, industrial facilities, forest areas, farmlands, restaurants, shopping centers, temples, roads, and bodies of water. Meteorological variables include temperature, relative humidity, surface pressure, boundary layer height, and meridional and zonal winds.

[16] A key issue of using satellite data to measure air quality in Taiwan is missing data due to cloud cover or high surface reflection (Liu et al. 2016, Tsai et al. 2011).

used in constructing the grid air quality data, see Jung et al. (2018) and Wang et al. (2020).

We collect data on atmospheric conditions, including daily average temperature, total rain hours, precipitation per hour, relative humidity, wind speed, and the number of hours that wind blows from each of the directions from the Taiwan Air Quality Monitoring Network, hosted by the Taiwan EPA. Panel B of table 1 reports the summary statistics of the key air pollutant, $PM_{2.5}$, and other weather variables included in the models. Figure 1 presents the average $PM_{2.5}$ concentration of each district and township in Taiwan during the study period. Air pollution is higher in the western coastal plain areas and is lower in the central and eastern mountainous areas.

## IV. Main Results

### A. *The Effects of Air Pollution on Accidents Caused by Risky Driving Behaviors*

Table 2 summarizes the relationship between air pollution and accidents caused by violations, using the fixed effects model specified in equation (1).[17] In model 1 we regress the number of accidents on $PM_{2.5}$ concentration without controlling for any weather conditions or fixed effects, and find a 1 $\mu g/m^3$ increase in the daily average of $PM_{2.5}$ concentration is associated with a 0.92% increase in the number of accidents. We add weather controls, including the linear and squared terms of daily average temperature, relative humidity, rainfall per hour, hours of rain, and wind speed in model 2, and still find a significantly positive association between the number of accidents and air pollution. As noted, these two models likely suffer from serious omitted variable biases, because region-specific factors, especially traffic volume and population, can affect both accidents caused by violations and air pollution.

In model 3, after region (district/township), year-by-month, and day-of-week fixed effects being controlled for, we find a *negative* association, significant at the 0.1% level, between air pollution and the number of accidents caused by violations. We deem that such a sign flip, after adding the fixed effects, is a plausible outcome. A region with a higher population is likely to have more traffic, traffic violations, accidents caused by violations, and a higher level of air pollution.[18] Failure to control for region fixed effects will therefore bias the effect of air pollution on risky driving behaviors upward. Similarly, higher traffic volume is potentially associated with more traffic violations and worse air quality. Omitting those time-varying factors will again bias the estimate upward. Still, this model may suffer from region-specific seasonal (year-by-month) and periodic (day of week) correlations between air pollution and traffic patterns. Therefore, in model 4 we instead include region-by-year-by-month and region-by-day-of-week fixed effects, and find a 1 $\mu g/m^3$ increase in the daily average of $PM_{2.5}$ is significantly associated with a 0.09% *decrease* in the number of accidents caused by violations. As a robustness check, we conduct an OLS estimation in model 5. We again find a significantly negative association, and the estimate is quantitatively similar. Indicated by the implied semi-elasticity, a 1 $\mu g/m^3$ increase in $PM_{2.5}$ is associated with a 0.12% decrease in the number of accidents caused by risky driving behaviors.

---

[17] Table A1 in the appendix presents the complete results, including the estimates of weather covariates.

To further address other potential endogeneity issues, e.g., from any remaining unobserved (region-specific) time-varying factors, reverse causality, and avoidance behaviors, we apply an instrumental variable approach with a control function. The first stage results are reported in table 3.[19] Regardless of the set of included instruments, the cluster-robust Kleibergen-Paap F statistics show that wind directions are relevant instruments for $PM_{2.5}$ concentration. We also find the effects of each wind direction are largely in line with the atmospheric evidence. For example, north winds move air pollutants to the downwind areas in the Southern air quality zone; and, with the mountain ranges stretching from north to south and slightly aligning to the east, west winds lead to higher levels of air pollution in the leeward-side Northern, Central, and Southern air quality zones.

---

[18] Indeed, by regressing the number of accidents on population density, we find a 100-person increase in population density per $km^2$ is associated with a 0.71% increase in the number of accidents caused by violations.

[19] Table A2 in the appendix presents the complete first-stage results, including the estimates of weather covariates.

We present the results using our IV strategy in table 4.[20] Model 1 does not include weather controls or fixed effects. With the air pollution being instrumented, we do not find any positive association between air pollution and accidents caused by driver violations. Still, because wind directions can affect many weather conditions such as temperature, precipitation, and humidity, violating the exclusion restriction is certainly of our concern.

Models 2 and 3, arguably the most credible models for identifying the casual relationships of our interests, yield virtually identical results. We find a 1 $\mu g/m^3$ increase in $PM_{2.5}$ is associated with a 0.56 – 0.59% *decrease* in the number of accidents caused by risky driving behaviors, and the estimates are both statistically significant at the 0.1% level. The K-P F-statistics show that wind directions are relevant instruments. The exclusion restriction is likely to hold after controlling for a host of weather covariates. Monotonicity is less of a concern when we allow the marginal effects of wind directions on air pollutions to be air quality zone-specific. We also estimate the second stage with OLS in model 4, and the implied semi-elasticity, -0.055, assures a robust result.

We thus far focus on the contemporaneous effect of air pollution on risky road behavior. Studies have also found that some health outcomes response to cumulative exposure to air pollution (Tsai et al. 2019). In model 5 we explore the cumulative (or lagged) effects of air pollution by including $PM_{2.5}$ concentrations in the three previous days. The results show that only same-day air pollution has impacts on accidents caused by drivers' risky behaviors. In the subsequent analyses, we use the specification of model 3 (PPML with IV, weather controls, and region-specific temporal fixed effects) to conduct several robustness tests as well as to explore the nonlinear treatment effects and transmission channels through which air pollution alters risky road behaviors.

---

[20] Table A3 in the appendix presents the complete estimation results, including the estimates of weather covariates.

### B. Robustness Checks

We argue that our IV strategy has adequately addressed our primary identification issue stemming from not including traffic volume. If the impacts of traffic volume had yet been treated by our IV strategy, the effects of air pollution on accidents caused by drivers' risky behaviors are likely heterogeneous between time and locations with different traffic volumes or patterns. For example, studies have documented that elevated air pollution decreases outdoor activities (e.g., Graff Zivin and Neidell 2009, Janke 2014, Saberian et al. 2017) and traffic from discretionary trips (Cutter and Neidell 2009). If the effects of such avoidance behaviors on (non-commuting) traffic volumes were still not accounted for by our instrument, we would expect to observe more negative effects of air pollution on accidents caused by violations during times with higher shares of discretionary trips (e.g., non-rush hours and weekends). To show that traffic volume no longer poses a threat to our causal interpretations, we examine whether the effects of air pollution are different for accidents that occurred in rush vs. non-rush hours, on weekdays vs. weekends, at intersections or not, as well as in regions with different population densities.

Table 5 reports the results. Models 1 and 2 use the daily number of accidents occurred during rush hours and non-rush hours as the outcome variable, respectively. Model 3 (model 4) uses the sample from weekdays (weekends) only. Models 5 and 6 use the number of accidents that occurred at an intersection and those not at an intersection, respectively. Model 7 (model 8) includes only samples from regions with population densities lower than (higher than or equal to) the sample median. We find the coefficients are all negative and similar in terms of magnitude. The results of model 1 to 4 suggest that the negative effects of air pollution on accidents caused by violations are unlikely explained by avoidance behaviors against air pollution. Indeed, we find the negative effect in rush hours is even larger in magnitude than that in non-rush hours (-0.0068 vs. -0.0052). Therefore, we conclude that our findings are not sensitive to, nor driven by, any remaining biases from omitted traffic volume-related variables.[21]

We have demonstrated a pathway to infer the positive association between the level of air pollution and individuals' degree of risk aversion. Increased risk aversion and less risky driving behaviors should have smaller effects on those accidents that are caused by mindless errors such as failing to judge another vehicle's path or speed. In model 9 we estimate the effect of air pollution on the number of accidents caused by errors, according to the primary contributory factor and the classification proposed by Reason et al. (1990). The effect, valued at -0.32%, is about half of that on accidents caused by violations (-0.59% in model 3 in table 4) and is imprecisely estimated. This result provides corroborating evidence for linking air pollution to the degree of risk aversion.[22]

In table A5 in the appendix, we report the estimation results using different specifications: (1) including only the linear terms for the weather controls, (2) including eight separate wind directions into as instruments, (3) including wind directions and speed as excluded instruments,[23] and (4) with daily observations aggregated at the spatially coarser county level to account for the effect of a driver's exposure to air pollution in regions other than the region where the accident occurred. In addition, we estimate a model including only accidents occurred on sunny days with good visibility, to minimize any potential effect from impaired visibility. All results are quantitatively similar and support our key finding: elevated air pollution decreases the number of accidents caused by driver violations.

---

[21] We estimate several models to test if elevated air pollution would reduce traffic volumes using daily data from Taipei City, the capital city of Taiwan. We do not find any evidence suggesting that air pollution would decrease traffic volume. These results again indicate that avoidance behavior cannot explain the reduction in traffic accidents cause by violations. The estimation results are presented in table A4 in the appendix.

[22] We should note that we classified accidents being caused by mindless errors based on the contributory factors listed in the accident reports conducted by polices. It is possible that some of these accidents are still caused by intended aggressive driving behaviors, but that such behaviors are not caught by the polices. Therefore, reducing risky driving behaviors can still reduce the number of accidents caused by errors, and this test is not completely a falsification exercise.

[23] Although wind speed has been used as an instrumental variable for air quality because of its relevance (e.g., He et al. 2019), we believe that wind speed is not a valid instrument in our case because studies have documented its effect on road safety (Dastoorpoor et al. 2016, Usman et al. 2012).

Lastly, we acknowledge that our estimates still include the effects of air pollution on accidents through other mechanisms such as impaired cognition, irritated respiratory systems, or increased aggression.[24] While a driver acts recklessly, impaired cognition or irritated respiratory organs due to elevated air pollution can increase the likelihood of an accident. Therefore, the effect of air pollution on reducing accidents through increased risk aversion is likely to be even larger when other effects are isolated. We more formally address this issue in the nonlinear effect of air pollution later.

## V. Transmission Channels, Nonlinearity, and Heterogenous Treatment Effects

### A. *The Transmission Channels of Air Pollution Affecting Risky Driving Behaviors*

Air pollution can affect mood and behaviors, including increased risk aversion, through both respiratory and visual channels. To shed light on the transmission channels through which air pollution affects driver behaviors and the resulting accidents, we explore the heterogeneous effects of air pollution (1) on drivers with different levels of direct exposure to air pollution and (2) under conditions whether air quality can be visually assessed. In light of our key findings and the credible evidence showing that non-enclosed vehicle drivers should on average have a much higher level of exposure to air pollution through their respiratory systems than drivers of enclosed vehicle do, we first hypothesize that air pollution would reduce the number of accidents primarily caused by violations committed by non-enclosed vehicle drivers more than those by enclosed vehicle drivers.[25]

---

[24] The mechanism of increased aggression is suggested by a few recent studies on the positive association between air pollution and crime (Bondy et al. 2020, Burkhardt et al. 2019, Herrnstadt et al. forthcoming).

Models 1 and 2 of table 6 present the results of the effects of air pollution on the number of accidents caused by violations committed by drivers of enclosed and non-enclosed vehicles, respectively.[26] The estimated effect on accidents caused by violations of non-enclosed vehicle drivers is only slightly more negative than that of enclosed vehicle drivers (-0.67% vs. -0.57%). Therefore, we find little evidence to show that air pollution affects road users' risky behaviors through the respiratory system.[27]

Subsequently, we explore whether air quality can affect risky driving behaviors through visual channels. As noted, it is much easier to assess air quality when there is natural light (Hyslop 2009, Yu et al. 2018). Therefore, we hypothesize that the effect would be stronger during times with ambient natural light than during times without. Models 3 and 4 in table 6 present the estimation results. We find that a 1 $\mu g/m^3$ increase in $PM_{2.5}$ leads to a 0.79% decrease in the number of accidents caused by risky driving behaviors when there is natural light. However, the effect of a 1 $\mu g/m^3$ increase in $PM_{2.5}$ is much smaller (-0.18%) and is imprecisely estimated in time without natural light. As a placebo test, we use the same IV specification to examine the effect of ozone, which is generally found to have little impact on the haziness of skies (Liu et al. 2020), on accidents caused violations. The results show an insignificant and slightly positive effect of ozone on accidents (see table A6 in the appendix). We therefore find strong evidence to demonstrate that air quality can affect risky driving behaviors through visual channels and infer that a person's degree of risk aversion can increase simply by "seeing" air pollution (i.e. when the skies are hazy).

---

[25] Sager (2019) conducted a similar investigation and expected that air pollution would have a stronger effect on increasing the number of non-enclosed vehicles involved in accidents than on enclosed vehicles. However, he finds that the positive effect is smaller on non-enclosed vehicles but provides no explanation. Our proposed hypothesis can therefore provide a coherent justification for this particularly puzzled result in Sager's study.

[26] We exclude the very small number of accidents principally caused by pedestrians' violations.

[27] It is possible that non-enclosed vehicle drivers perform defense health behaviors such as wearing masks. However, we are not aware of any evidence indicating a large share of non-enclosed vehicle drivers regularly wearing masks that can effectively filter out $PM_{2.5}$, the air pollutant that our study focuses on. On the other hand, enclosed vehicle drivers, if they keep the windows open, might still expose to similar levels of air pollution to those among non-enclosed vehicle drivers. However, with nearly all enclosed vehicles equipped with air condition and the hot and rainy subtropical climate in Taiwan, it is uncommon for enclosed vehicle drivers to have their windows open, especially in populated areas. We therefore believe that these unobserved behaviors are unlikely to close the gap in direct air pollution exposure between enclosed and non-enclosed vehicle drivers, which has been well documented in the literature (Dirks et al. 2012, Rivas et al. 2017, Tsai et al. 2008, Vlachokostas et al. 2012).

### B. *Nonlinear and Heterogeneous Treatment Effects*

Medical literature has shown that the dose-response relationships between air pollution and health outcomes can be nonlinear (e.g., Daniels et al. 2000, Dominici et al. 2002). Several economic studies have also found nonlinear effects of air pollution on various societal outcomes, e.g., Burkhardt et al. (2019) on crime, Ebenstein et al. (2016) on life expectancy, and Schlenker and Walker (2016) on hospitalization costs. We are not aware of any exploration of the nonlinearity in the effect of air pollution on risk attitude or risky behaviors. We subsequently examine whether such nonlinearity exists.

Based on each region's average $PM_{2.5}$ concentration, we stratify all regions into two groups (the better and worse 50) using the sample median of daily average $PM_{2.5}$ concentration of the entire study period as the cutoff. Models 1 and 2 in table 7 present the estimation results from the models using these two subsamples, respectively. We find the negative effect of air pollution on the number of accidents caused by violations is only significant in the worse 50 regions with higher level of air pollution, and the effect size is more than doubled of that in the regions with better air quality (-0.62% vs. -0.26%). The average $PM_{2.5}$ concentration in the data of Sager (2019) is 13.32 $\mu g/m^3$, and he finds a 1 $\mu g/m^3$ increase in $PM_{2.5}$ is associated with 0.3 – 0.6% more accidents. The average PM2.5 concentration in our top and bottom halves of the regions are 20.42 and 33.56 $\mu g/m^3$, respectively. Interestingly, by combining the two estimates from our study and that from Sager (2019), we observe the effect of air pollution on road accidents monotonically decreases with higher level of air pollution.

We further leverage control function's ability to deal with the nonlinearity in the endogenous variable in the second stage and include linear splines of $PM_{2.5}$ concentration in model 3 (Henderson and Souto 2018, Wooldridge 2015). We find a positive, albeit insignificant, effect of air pollution on accidents when the $PM_{2.5}$ concentration is below or equal to 10 $\mu g/m^3$, and only identify the negative effects when the $PM_{2.5}$ concentration is above 10 $\mu g/m^3$. This result provides corroborating evidence of the nonlinearity in the effect of air pollution on risky road behaviors.

These results suggest that the marginal effect of air pollution on reducing accidents through increased risk aversion rises more quickly than the effect on increasing accidents through impaired cognition (or irritated respiratory systems) with elevated air pollution levels. We illustrate such relationships conceptually in figure 2. The dotted line shows the total (accumulated) effect size of air pollution on *decreasing* accidents through increased levels of risk aversion, and the dashed line shows the effect on *increasing* accidents through cognitive impairment. The solid line represents the combined marginal effect, which is the difference between the accumulated effect from increased risk aversion and that from cognitive impairment. Our data and analyses have recovered the marginal effect when air pollution is relatively high, for example, the shaded area (I), where the size of the accumulated effect of increased risk aversion is larger than that of cognitive impairment. The results in Sager (2019), on the other hand, may have traced out the positive marginal effect of air pollution on accidents in area (II).[28] This illustration not only highlights the nonlinear effect of air pollution on risky aversion or impaired cognition, but shows that our results can actually be "consistent" with those from Sager.[29]

Many studies have found that gender and age are two of the key characteristics for explaining an individual's risk tolerance (e.g., Borghans et al. 2009, Halek and Eisenhauer 2001, Hartog et al. 2002). We examine whether the effects of air pollution on risky road behaviors are heterogeneous across gender and age groups. Table 8 presents the associated results. In models 1 and 2, the outcome variables are the number of accidents caused by violations for which a female or male driver is primarily liable. The effects are virtually identical. Therefore, we do not find evidence to show that air pollution heterogeneously influences drivers of different biological gender.

In models 3 to 5, we use the counts of the most liable driver in an accident by age groups (below 40, 40 to 64, and 65 and above) as the outcome variables. We find the effects to be largely homogeneous across the two younger groups (-0.61% and -0.62%). However, the effect on the number of accidents primarily caused by violations committed by drivers 65 years old and above is much lower (-0.34%) and imprecisely estimated.[30] Corroborating with the results showing nonlinear effects, this heterogeneity suggests that the effect on cognition may be relatively stronger, compared to the effect on risk aversion, among elderly drivers. This can be illustrated by an upward shift of the effect of air pollution on cognition (the dashed line) in figure 2, while holding the effect of air pollution on risk aversion (the dotted line).[31]

---

[28] Moreover, our model that includes the linear splines of air pollution, model (3) in table 7, provides indicative evidence of the shape of the entire solid curve. Despite the insignificant positive effect when $PM_{2.5}$ concentration is below 10 $\mu g/m^3$, the model suggests that the net marginal effect equals to zero when $PM_{2.5}$ concentration is 18.22 $\mu g/m^3$, which is in line with the estimates at different average levels of air pollution from our study (model 1 and 2 in table 7) and those from Sager (2019). Still, we note that the estimates of ours and those of Sager (2019) are not necessarily comparable, given that we primarily focus on accidents caused on violations and the traffic conditions and dynamics are likely very different between Taiwan and the U.K.

[29] Although our example depicts both of the dose-response relationships being nonlinear, we can recover a similar combined marginal effect of air pollution on accidents as long as one of the dose-responses is nonlinear, assuming there is no effect of air pollution on cognition nor risk aversion when the level of ai pollution is zero.

[30] Trips of those 65 years old and above are more likely discretionary. Therefore, the smaller negative effect of air pollution among elderlies once again suggest that avoidance behaviors against air pollution cannot explain the reduction in accidents.

## VI. Conclusion and Discussion

Despite a large strand of literature on the socio-economic costs of air pollution, in this study we find a rare "benefit" of air pollution: reducing accidents caused by risky behaviors on roads. Using administrative traffic accident records and high-resolution air quality data of Taiwan, we identify air quality as a new factor for changing life-threatening risky behaviors. Our results find that elevated air pollution, measured by $PM_{2.5}$ concentration, leads to fewer traffic accidents caused by driver violations, but has no significant effect on those caused by mindless errors. Specifically, we find that a 1 $\mu g/m^3$ increase in $PM_{2.5}$ concentration leads to a 0.59% decrease in accidents caused by driver violations. That is, a one-standard deviation increases in $PM_{2.5}$ concentration can decrease such accidents by more than 9%. In addition, the nonlinearity in the dose-response relationship between air pollution and accidents suggest that, among the effects of air pollution on risk aversion and cognition, at least one of them is nonlinear. These results in particular add to the literature on the determinants of risky behaviors (Dohmen et al. 2011) and (non)stability of risk preferences (Schildberg-Hörisch 2018)

Cognitive ability and risk tolerance have been shown to be interrelated in determining many economic outcomes such as investment in human capital and inequality (Dohmen et al. 2018, Heckman et al. 2006). Our results by no means lead us to advocate for less stringent air pollution regulation, nor do they suggest that the costs of air pollution have been overestimated. Instead, given that the effects of air pollution on risk aversion and impaired cognition (and irritated respiratory systems) can influence risky behaviors in the opposite directions, the estimated cost of air pollution on cognitive performance and other associated health outcomes involving risk attitudes may be underestimated, if the effect on risk attitudes is not isolated.[32]

---

[31] Elderlies are in general more risk averse (Halek and Eisenhauer 2001, Hartog et al. 2002), so another natural explanation for the heterogeneity across age groups is that air pollution has a smaller effect on risky behaviors among those who are more risk averse. However, since females are on average more risk averse, the homogeneous effects across genders do not support such explanation.

We further find such negative effects are only observed in times when air quality can be visually assessed, i.e., when natural light is available. However, the effect is largely homogeneous across the drivers of enclosed and non-enclosed vehicle, who arguably are exposed to very different levels of air pollution. These results show that air quality can affect risky driving behaviors through visual channels, but there is no evidence to validate the respiratory route. To the best of our knowledge, we are the first economic study that is able to simultaneously test these two transmission channels. This finding also implies that, when quantifying the effects of different air pollutants on societal outcomes, one should pay attention to how the focused pollutants can change an individual's visual perception of the environment. For example, ozone in general has small impact on visibility and can even makes the sky bluer, which is visually pleasant to many; therefore, ozone may simultaneously weaken cognitive performance and reduce risk aversion. Failure to isolate one effect from the other, again, may prejudice the casual interpretation of the estimated effects of ozone pollution.

---

[32] For example, He et al. (2019) find that air pollution decreases the productivity of call center workers, primarily through an increase in breaks. For some jobs, without precise measures of output, taking more breaks could be directly considered to reflect lower productivity and risks one's job security. A worker with a higher level of risk aversion might refrain from taking breaks. Such behavior would mitigate some observable losses in productivity due to elevated air pollution. On the other hand, if lower level of risk tolerance makes an individual who feels ill more likely to seek a consultation of a physician, failure to isolate the effect of risk preferences may lead to overestimation of the cost of air pollution measured by medical expenditure.

Some might see our key finding—elevated air pollution decreases accidents caused by risky driving behaviors—surprising and even in stark contrast to common wisdom like the results in Sager (2019), who used a similar research design but focused on *all* types of traffic accidents in the United Kingdom. Indeed, we find no significant effect of air pollution on traffic accidents caused by harmless lapses nor those caused by violations in areas with more comparable air quality to that in highly developed countries. Our results show a nonlinear dose-response relationship between air pollution and risky behaviors on roads. That is, air pollution likely increases the degree of risk aversion at an increasing rate or at least at a rate faster than that on reducing cognition. Corroborating our results with those from Sager (2019) enhances the external validity of the estimated effects of air pollution on road safety and risky behaviors in general. Our empirical results have more direct implications and are more readily transferable to many non-high-income countries, where the climates, environmental quality, and traffic compositions are more similar to those in Taiwan. Moreover, the nonlinearity in air pollution dose-response reinforces the motivation for investigating the distributional effects of air pollution across groups with different levels of exposure (Hsiang et al. 2019). Exploring the nonlinear effects of air pollution on different risky behaviors or direct measures of risk aversion is an interesting avenue for future work. Lastly, another straightforward extension will be to investigate the effects of air pollution on more common outcomes of risky driving behaviors such as traffic violations, especially speeding caught by enforcement cameras.

# References

Anderson, M. L. (2020). As the wind blows: The effects of long-term exposure to air pollution on mortality. *Journal of the European Economic Association*, *18*(4), 1886-1927.

Anstey, K. J., Horswill, M. S., Wood, J. M., & Hatherly, C. (2012). The role of cognitive and visual abilities as predictors in the multifactorial model of driving safety. *Accident Analysis & Prevention*, *45*, 766-774.

Arceo, E., Hanna, R., & Oliva, P. (2016). Does the effect of pollution on infant mortality differ between developing and developed countries? Evidence from Mexico City. *The Economic Journal*, *126*(591), 257-280.

Archsmith, J., Heyes, A., & Saberian, S. (2018). Air quality and error quantity: Pollution and performance in a high-skilled, quality-focused occupation. *Journal of the Association of Environmental and Resource Economists*, *5*(4), 827-863.

Bassi, A., Colacito, R., & Fulghieri, P. (2013). 'O sole mio: An experimental analysis of weather and risk attitudes in financial decisions. *The Review of Financial Studies*, *26*(7), 1824-1852.

Bedi, A. S., Nakaguma, M. Y., Restrepo, B. J., & Rieger, M. (2021). Particle pollution and cognition: Evidence from sensitive cognitive tests in Brazil. *Journal of the Association of Environmental and Resource Economists*, *8*(3), 443-474.

Blundell, R., & Powell, J. L. (2003). Endogeneity in nonparametric and semiparametric regression models. In *Advances in Economics and Econonometrics: Theory and Applications*, Eighth World Congress, Volume 2, ed. Mathias Dewatripont, Lars Hansen, and Stephen Turnovsky, 312–57. Cambridge: Cambridge University Press

Bondy, M., Roth, S., & Sager, L. (2020). Crime is in the air: The contemporaneous relationship between air pollution and crime. *Journal of the Association of Environmental and Resource Economists*, *7*(3), 555-585.

Borghans, L., Heckman, J. J., Golsteyn, B. H., & Meijers, H. (2009). Gender differences in risk aversion and ambiguity aversion. *Journal of the European Economic Association*, *7*(2-3), 649-658.

Brijs, T., Karlis, D., & Wets, G. (2008). Studying the effect of weather conditions on daily crash counts using a discrete time-series model. *Accident Analysis & Prevention*, *40*(3), 1180-1190.

Brimblecombe, P. (1987). *The Big Smoke: A History of Air Pollution in London since Medieval Times*. Routledge.25

*Research*, *19*(4), 896-910.

Middleton, P., Stewart, T. R., & Leary, J. (1985). On the use of human judgment and physical/chemical measurements in visual air quality management. *Journal of the Air Pollution Control Association*, *35*(1), 11-18.

Moretti, E., & Neidell, M. (2011). Pollution, health, and avoidance behavior evidence from the ports of Los Angeles. *Journal of Human Resources*, *46*(1), 154-175.

Neidell, M. (2009). Information, avoidance behavior, and health the effect of ozone on asthma hospitalizations. *Journal of Human Resources*, *44*(2), 450-478.

Ouimette, J. R., & Flagan, R. C. (1982). The extinction coefficient of multicomponent aerosols. *Atmospheric Environment (1967)*, *16*(10), 2405-2419.

Pasidis, I. (2017). Congestion by accident? A two-way relationship for highways in England. *Journal of Transport Geography*.

Persico, C. L., & Venator, J. (2021). The effects of local industrial pollution on students and schools. *Journal of Human Resources*, *56*(2), 406-445.

Quddus, M. A., Wang, C., & Ison, S. G. (2009). Road traffic congestion and crash severity: econometric analysis using ordered response models. *Journal of Transportation Engineering*, *136*(5), 424-435.

Reason, J., Manstead, A., Stradling, S., Baxter, J., & Campbell, K. (1990). Errors and violations on the roads: a real distinction? *Ergonomics*, *33*(10-11), 1315-1332.

Rivas, I., Kumar, P., & Hagen-Zanker, A. (2017). Exposure to air pollutants during commuting in London: are there inequalities among different socio-economic groups? *Environment International*, *101*, 143-157.

Rosenblitt, J. C., Soler, H., Johnson, S. E., & Quadagno, D. M. (2001). Sensation seeking and hormones in men and women: exploring the link. *Hormones and Behavior*, *40*(3), 396-402.

Saberian, S., Heyes, A., & Rivers, N. (2017). Alerts work! Air quality warnings and cycling. *Resource and Energy Economics*, *49*, 165-185.

Sager, L. (2019). Estimating the effect of air pollution on road safety using atmospheric temperature inversions. *Journal of Environmental Economics and Management*, *98*, 102250.

Saunders, E. M. (1993). Stock prices and Wall Street weather. *The American Economic Review*, *83*(5), 1337-1345.

Schildberg-Hörisch, H. (2018). Are risk preferences stable? *Journal of Economic Perspectives*, *32*(2), 135-54.30

## Tables

Table 1 Summary Statistics of Accident Counts and Weather Conditions

| Variable | Mean | Median | S.D. | Min | Max | Total |
|---|---|---|---|---|---|---|
| Panel A | | | | | | |
| Daily Accidents Caused by Violations | | | | | | |
|     All | 1.27 | 1 | 2.22 | 0 | 33 | 1,126,395 |
|     By non-enclosed vehicle drivers | 0.60 | 0 | 1.26 | 0 | 20 | 534,493 |
|     With natural light | 0.89 | 0 | 1.61 | 0 | 24 | 787,990 |
|     Fair-weather only | 1.04 | 0 | 1.99 | 0 | 29 | 926,417 |
|     During rush hours | 0.54 | 0 | 1.10 | 0 | 16 | 482,955 |
|     At intersections | 0.84 | 0 | 1.60 | 0 | 25 | 744,665 |
|     By female drivers | 0.41 | 0 | 0.93 | 0 | 15 | 366,026 |
| Panel B | | | | | | |
| Air Quality and Atmospheric Conditions | | | | | | |
|     $PM_{2.5}$ (μg/m$^3$) | 27.06 | 23.40 | 15.81 | 0.36 | 241.41 | |
|     Avg. Temp. (°C) | 21.76 | 22.86 | 5.11 | 4.36 | 31.32 | |
|     Rainfall (mm/hr) | 0.20 | 0.00 | 0.71 | 0.00 | 32.50 | |
|     Rain hours (hr) | 2.47 | 0.35 | 4.38 | 0.00 | 24.00 | |
|     Relative Humidity (%) | 81.37 | 81.92 | 7.94 | 44.59 | 99.60 | |
|     Wind Speed (m/s) | 2.23 | 1.91 | 1.13 | 0.32 | 16.11 | |
|     East Wind (%) | 29 | 24 | 23 | 0 | 100 | |
|     South Wind (%) | 19 | 12 | 20 | 0 | 100 | |
|     West Wind (%) | 25 | 24 | 18 | 0 | 100 | |
|     North Wind (%) | 27 | 21 | 24 | 0 | 100 | |

Notes: Mean, median, S.D., min, and max are based on 888,736 observations from each day of each of all 365 districts and townships between 2009 – 2015.

Table 2. Air Pollution and Traffic Accidents Caused by Violations (Fixed Effects Models)

| Variables | (1) PPML | (2) PPML | (3) PPML | (4) PPML | (5) OLS |
|---|---|---|---|---|---|
| $PM_{2.5}$ | 0.0092*** | 0.0120*** | -0.0010*** | -0.0009*** | -0.0012*** |
|  | (0.0016) | (0.0021) | (0.0002) | (0.0001) | (0.0002) |
| Semi-Elasticity |  |  |  |  | -0.0009 |
|  |  |  |  |  |  |
| Observations | 892,044 | 888,736 | 886,182 | 820,358 | 888,736 |
| Weather Controls | N | Y | Y | Y | Y |
| Town FE | N | N | Y | N | N |
| Year-Month FE | N | N | Y | N | N |
| DoW FE | N | N | Y | N | N |
| Town-Year-Month FE | N | N | N | Y | Y |
| Town-DoW FE | N | N | N | Y | Y |
| Pseudo $R^2$ / Adj-$R^2$ | 0.008 | 0.015 | 0.468 | 0.465 | 0.734 |

Notes: The outcome variable is the total number of accidents caused by violations in each district/township. Weather controls include the linear and squared terms of daily average temperature, relative humidity, per hour rainfall, hours of rain, and wind speed. Standard errors are in parenthesis and clustered at number of day and district/township level. *, **, ***: significant at 5%, 1%, and 0.1%

Table 3. First Stage Results of Air Pollution on Wind Directions

|  | (1) | (2) | (3) |
|---|---|---|---|
| Variables (WD by AQ Zone) | OLS | OLS | OLS |
| South Wind x AQ Zone North | 0.7096 | 3.7565** | -0.3776 |
|  | (1.9483) | (1.3526) | (1.4463) |
| South Wind x AQ Zone Central | -2.7817* | -3.3785*** | -3.1097*** |
|  | (1.2435) | (0.7933) | (0.8555) |
| South Wind x AQ Zone South | -10.1895*** | -4.1085*** | 0.2931 |
|  | (0.9363) | (0.9309) | (1.0165) |
| South Wind x AQ Zone East | -12.1673*** | -0.5219 | -5.8214*** |
|  | (1.2571) | (0.9196) | (0.7752) |
| West Wind x AQ Zone North | 3.6997** | 6.2390*** | 4.7087*** |
|  | (1.1534) | (0.7854) | (0.8303) |
| West Wind x AQ Zone Central | 8.1869** | 3.1903*** | 2.4498* |
|  | (2.9403) | (0.7910) | (0.9891) |
| West Wind x AQ Zone South | 15.7092*** | 2.1208* | 4.9465*** |
|  | (1.4134) | (0.9456) | (1.0416) |
| West Wind x AQ Zone East | -1.7700 | -7.4636*** | -5.7200*** |
|  | (1.3314) | (0.9962) | (1.0923) |
| North Wind x AQ Zone North | 1.0580 | 3.6321** | 1.5687 |
|  | (1.0308) | (1.1696) | (1.3506) |
| North Wind x AQ Zone Central | 16.9963*** | -1.6744* | -5.2331*** |
|  | (1.3061) | (0.6817) | (0.8904) |
| North Wind x AQ Zone South | 26.7380*** | 10.0435*** | 9.0299*** |
|  | (1.6625) | (1.0584) | (1.1472) |
| North Wind x AQ Zone East | -4.5505* | -4.1387*** | -3.2284** |
|  | (1.8290) | (1.0475) | (1.0639) |
|  |  |  |  |
| Observations | 892,044 | 888,736 | 888,736 |
| Weather Controls | N | Y | Y |
| Town FE | N | Y | N |
| Year-Month FE | N | Y | N |
| DoW FE | N | Y | N |
| Town-Year-Month FE | N | N | Y |
| Town-DoW FE | N | N | Y |
| K-P F-statistics | 99.67 | 46.27 | 31.31 |
| R-squared | 0.221 | 0.617 | 0.656 |

Notes: The outcome variable is the daily average of $PM_{2.5}$ concentration in each district/township. The estimates represent the air quality zone specific marginal effect of an additional hour that wind blows from a certain direction on $PM_{2.5}$ concentration. Weather controls include the linear and squared terms of daily average temperature, relative humidity, per hour rainfall, hours of rain, and wind speed. Standard errors are in parenthesis and clustered at number of day and district/township level.

*, **, ***: significant at 5%, 1%, and 0.1%

Table 4. Air Pollution and Traffic Accidents Caused by Violations (IV and Control Function)

| Variables | (1) IV PPML | (2) IV PPML | (3) IV PPML | (4) IV OLS | (5) IV PPML |
|---|---|---|---|---|---|
| $PM_{2.5}$ | -0.0013 | -0.0056*** | -0.0059*** | -0.0070*** | -0.0059*** |
|  | (0.0065) | (0.0012) | (0.0012) | (0.0016) | (0.0012) |
| Semi-Elasticity |  |  |  | -0.0055 |  |
| $PM_{2.5}$ (1 day before) |  |  |  |  | 0.0001 |
|  |  |  |  |  | (0.0002) |
| $PM_{2.5}$ (2 days before) |  |  |  |  | 0.0000 |
|  |  |  |  |  | (0.0002) |
| $PM_{2.5}$ (3 days before) |  |  |  |  | 0.0004* |
|  |  |  |  |  | (0.0002) |
| Observations | 892,044 | 886,182 | 820,358 | 888,736 | 820,355 |
| Weather Controls | N | Y | Y | Y | Y |
| Town FE | N | Y | N | N | N |
| Year-Month FE | N | Y | N | N | N |
| DoW FE | N | Y | N | N | N |
| Town-Year-Month FE | N | N | Y | Y | Y |
| Town-DoW FE | N | N | Y | Y | Y |
| K-P F-statistics | 99.67 | 46.27 | 31.31 | 31.31 | 31.31 |

Notes: The outcome variable is the total number of accidents caused by violations in each district/township. Weather controls include the linear and squared terms of daily average temperature, relative humidity, per hour rainfall, hours of rain, and wind speed. Standard errors are in parenthesis and clustered at number of day and district/township level. *, **, ***: significant at 5%, 1%, and 0.1%

Table 5. Robustness Checks: Heterogenous Effects across Traffic Patterns, Lagged Effects, and Effect on Errors

| Variables | (1) Rush Hours | (2) Non-rush Hours | (3) Weekdays | (4) Weekend | (5) At Intersections | (6) Not at Intersections | (7) Regions w/ Low Pop | (8) Regions w/ High Pop | (9) Mindless Errors |
|---|---|---|---|---|---|---|---|---|---|
| $PM_{2.5}$ | -0.0068*** | -0.0052*** | -0.0060*** | -0.0060** | -0.0054*** | -0.0070*** | -0.0064* | -0.0057*** | -0.0032 |
|  | (0.0017) | (0.0013) | (0.0014) | (0.0023) | (0.0013) | (0.0018) | (0.0032) | (0.0013) | (0.0019) |
| Observations | 763,893 | 789,956 | 568,466 | 211,482 | 748,712 | 782,263 | 375,139 | 445,219 | 776,785 |

Notes: The outcome variable is the associated number of accidents in each district/township in model 1 to 8, and that caused by errors in model 9. All models control for Town-Year-Month and Town-DoW fixed effects as well as weather controls including the linear and squared terms of daily average temperature, relative humidity, per hour rainfall, hours of rain, and wind speed. $PM_{2.5}$ concentration is instrumented by wind directions. Standard errors are in parenthesis and clustered at number of day and district/township level.

*, **, ***: significant at 5%, 1%, and 0.1%

Table 6. Transmission Channels: Heterogeneous Effects by Driver Type and Light Conditions

| Variables | (1) Enclosed Vehicle | (2) Non-enclosed Vehicle | (3) Natural Light | (4) No Natural Light |
|---|---|---|---|---|
| PM$_{2.5}$ | -0.0057*** | -0.0067*** | -0.0079*** | -0.0018 |
|  | (0.0016) | (0.0016) | (0.0015) | (0.0018) |
|  |  |  |  |  |
| Observations | 781,092 | 762,527 | 807,090 | 698,196 |

Notes: The outcome variable is the associated number of accidents caused by violations in each district/township. All models control for Town-Year-Month and Town-DoW fixed effects as well as weather controls including the linear and squared terms of daily average temperature, relative humidity, per hour rainfall, hours of rain, and wind speed. PM$_{2.5}$ concentration is instrumented by wind directions. Standard errors are in parenthesis and clustered at number of day and district/township level.

*, **, ***: significant at 5%, 1%, and 0.1%

Table 7. Nonlinear Effect of Air Pollution on Traffic Accidents Caused by Violations

| Variables | (1) Better 50 | (2) Worse 50 | (3) PM$_{2.5}$ Splines |
|---|---|---|---|
| PM$_{2.5}$ | -0.0026 (0.0025) | -0.0062*** (0.0014) | |
| PM$_{2.5}$ (0 – 10 μg/m$^3$) | | | 0.0060 (0.0046) |
| PM$_{2.5}$ (10 – 20 μg/m$^3$) | | | -0.0073*** (0.0015) |
| PM$_{2.5}$ (20 – 30 μg/m$^3$) | | | -0.0063*** (0.0013) |
| PM$_{2.5}$ (30 μg/m$^3$ and above) | | | -0.0054*** (0.0012) |
| Observations | 378,385 | 441,973 | 820,358 |

Notes: The outcome variable is the total number of accidents caused by violations in each district/township. All models control for Town-Year-Month and Town-DoW fixed effects as well as weather controls including the linear and squared terms of daily average temperature, relative humidity, per hour rainfall, hours of rain, and wind speed. PM$_{2.5}$ concentration is instrumented by wind directions. Standard errors are in parenthesis and clustered at number of day and district/township level.

*, **, ***: significant at 5%, 1%, and 0.1%

Table 8. Heterogeneous Effects by Gender and Age

| Variables | (1) Female | (2) Male | (3) Under 40 | (4) 40 to 64 | (5) 65 and Above |
|---|---|---|---|---|---|
| $PM_{2.5}$ | -0.0058** | -0.0059*** | -0.0061*** | -0.0062** | -0.0034 |
| | (0.0018) | (0.0013) | (0.0017) | (0.0019) | (0.0031) |
| Observations | 715,911 | 806,257 | 659,313 | 651,161 | 541,750 |

Notes: The outcome variable is the associated number of accidents caused by violations in each district/township. All models control for Town-Year-Month and Town-DoW fixed effects as well as weather controls including the linear and squared terms of daily average temperature, relative humidity, per hour rainfall, hours of rain, and wind speed. $PM_{2.5}$ concentration is instrumented by wind directions. Standard errors are in parenthesis and clustered at number of day and district/township level.

*, **, ***: significant at 5%, 1%, and 0.1%

**Figures**

Figure 1. PM$_{2.5}$ Concentration by District/Township in Taiwan, 2009 – 2015

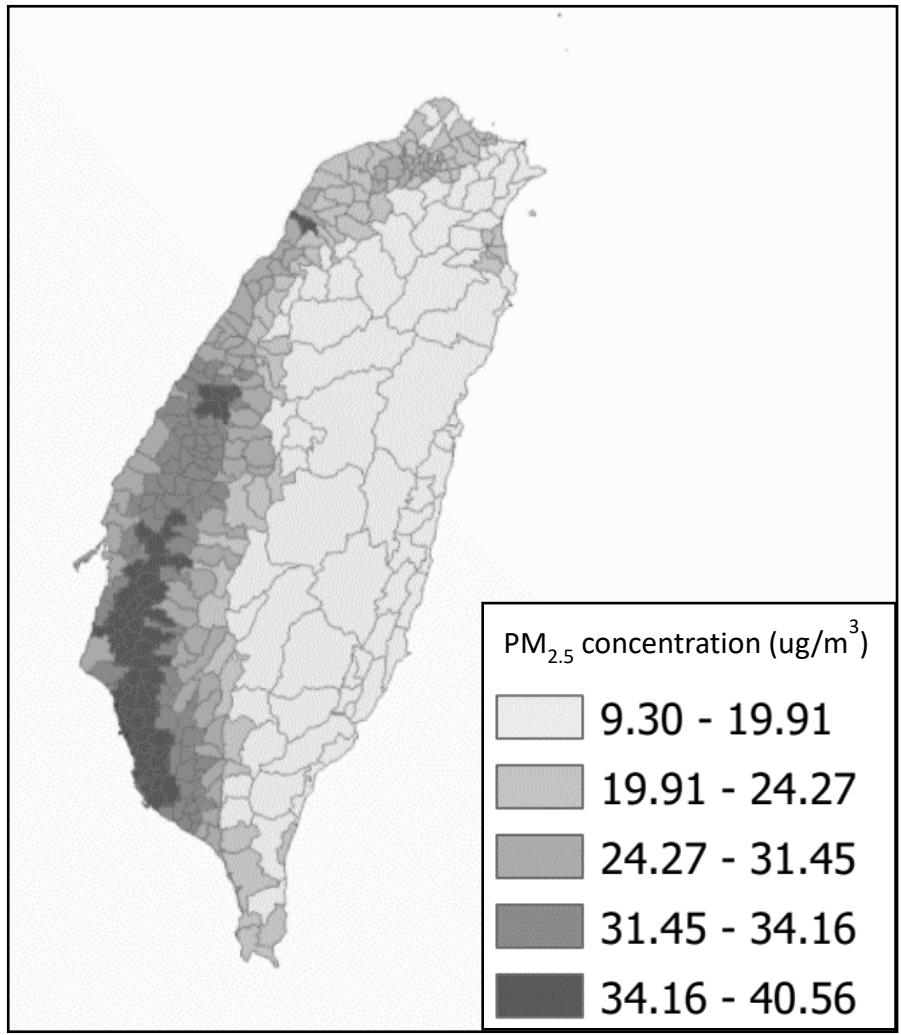

Notes: The cutoffs are 20, 40, 60, 80 centiles, respectively.

Figure 2. Nonlinear Effects of Air Pollution on Cognition, Risk Aversion, and Accidents

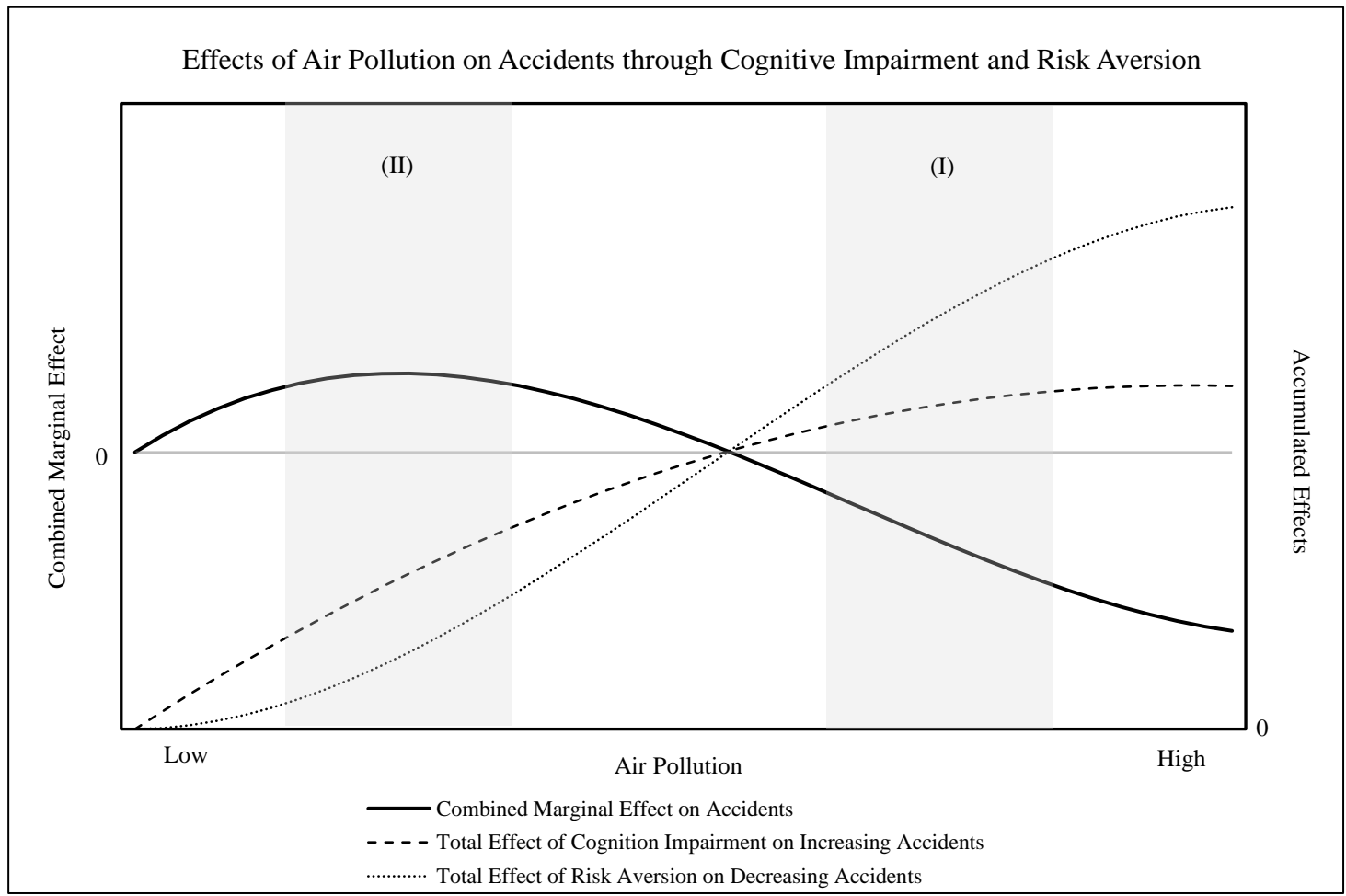

Notes: The magnitude of the combined marginal effect of air pollution on traffic accidents is shown on the primary axis on the left. The horizontal reference line in the middle of the figure indicates zero marginal effect. The absolute total (accumulated) effects of air pollution on traffic accidents through impairment cognition and increased risk aversion are on the secondary (right) axis. The magnitude of the total effects begins at zero from the bottom of the figure.

# Appendix

Table A1. Air Pollution and Traffic Accidents Caused by Violations (Fixed Effects Models)

| Variables | (1) PPML Coefficient | SE | (2) PPML Coefficient | SE | (3) PPML Coefficient | SE | (4) PPML Coefficient | SE | (5) OLS Coefficient | SE |
|---|---|---|---|---|---|---|---|---|---|---|
| $PM_{2.5}$ | 0.0092*** | (0.0016) | 0.0120*** | (0.0021) | -0.0010*** | (0.0002) | -0.0009*** | (0.0001) | -0.0012*** | (0.0002) |
| Temp. (°C) | | | -0.1660*** | (0.0178) | -0.0068 | (0.0042) | -0.0097* | (0.0046) | | |
| [Temp. (°C)]$^2$ | | | 0.0044*** | (0.0005) | 0.0003** | (0.0001) | 0.0004** | (0.0001) | | |
| Rainfall (mm/hr) | | | -0.0325 | (0.0372) | -0.0212** | (0.0070) | -0.0195** | (0.0069) | | |
| [Rainfall (mm/hr)]$^2$ | | | 0.0009 | (0.0024) | -0.0015 | (0.0009) | -0.0016 | (0.0009) | | |
| Rain hours (hr) | | | 0.0008 | (0.0073) | -0.0023 | (0.0013) | -0.0016 | (0.0013) | | |
| [Rain hours (hr)]$^2$ | | | -0.0000 | (0.0002) | -0.0001 | (0.0001) | -0.0001 | (0.0001) | | |
| Relative Humidity (%) | | | -0.0548** | (0.0171) | -0.0001 | (0.0037) | -0.0028 | (0.0035) | | |
| [Relative Humidity (%)]$^2$ | | | 0.0004*** | (0.0001) | -0.0000 | (0.0000) | -0.0000 | (0.0000) | | |
| Wind Speed (m/s) | | | 0.0868 | (0.0964) | 0.0493*** | (0.0085) | 0.0355*** | (0.0073) | | |
| [Wind Speed (m/s)]$^2$ | | | -0.0220 | (0.0117) | -0.0061*** | (0.0014) | -0.0042*** | (0.0013) | | |
| Constant | -0.0247 | (0.0844) | 3.2903*** | (0.7548) | 1.0937*** | (0.1533) | 1.2706*** | (0.1471) | 1.8061*** | (0.2171) |
| Observations | 892,044 | | 888,736 | | 886,182 | | 820,358 | | 888,736 | |
| Weather Controls | N | | Y | | Y | | Y | | Y | |
| Town FE | N | | N | | Y | | N | | N | |
| Year-Month FE | N | | N | | Y | | N | | N | |
| DoW FE | N | | N | | Y | | N | | N | |
| Town-Year-Month FE | N | | N | | N | | Y | | Y | |
| Town-DoW FE | N | | N | | N | | Y | | Y | |
| Pseudo $R^2$ / Adj-$R^2$ | 0.00759 | | 0.0147 | | 0.468 | | 0.465 | | 0.734 | |

Notes: This table presents complete results of table 2. The outcome variable is the total number of accidents caused by violations in each district/township. Standard errors are clustered at number of day and district/township level.

*, **, ***: significant at 5%, 1%, and 0.1%

Table A2. First Stage Results of Air Pollution on Wind Directions and Weather Conditions

| Variables | (1) PPML Coefficient | SE | (2) PPML Coefficient | SE | (3) PPML Coefficient | SE |
|---|---|---|---|---|---|---|
| Temp. (°C) | | | 2.6814*** | (0.2536) | 2.2166*** | (0.3431) |
| [Temp. (°C)]² | | | -0.0669*** | (0.0063) | -0.0602*** | (0.0084) |
| Rainfall (mm/hr) | | | -0.3305 | (0.1764) | -0.2096 | (0.1837) |
| [Rainfall (mm/hr)]² | | | 0.0311* | (0.0128) | 0.0288* | (0.0135) |
| Rain hours (hr) | | | -0.4626*** | (0.0619) | -0.5582*** | (0.0644) |
| [Rain hours (hr)]² | | | 0.0165*** | (0.0030) | 0.0183*** | (0.0031) |
| Relative Humidity (%) | | | 0.7863** | (0.2703) | 1.2838*** | (0.2797) |
| [Relative Humidity (%)]² | | | -0.0075*** | (0.0017) | -0.0101*** | (0.0017) |
| Wind Speed (m/s) | | | -4.5707*** | (0.3673) | -5.4300*** | (0.3927) |
| [Wind Speed (m/s)]² | | | 0.2979*** | (0.0426) | 0.3675*** | (0.0440) |
| Constant | 20.8697*** | (0.5675) | -3.6722 | (11.3294) | -18.4489 | (11.9665) |
| | | | | | | |
| Observations | 892,044 | | 888,736 | | 888,736 | |
| Weather Controls | N | | Y | | Y | |
| Town FE | N | | Y | | N | |
| Year-Month FE | N | | Y | | N | |
| DoW FE | N | | Y | | N | |
| Town-Year-Month FE | N | | N | | Y | |
| Town-DoW FE | N | | N | | Y | |
| K-P F-statistics | 99.67 | | 46.27 | | 31.31 | |
| R-squared | 0.221 | | 0.617 | | 0.656 | |

Notes: This table presents complete results of table 3. The outcome variable is the daily average of PM2.5 concentration in each district/township. Standard errors are clustered at number of day and district/township level. See table 3 for the estimates of wind directions by each air quality zone.
*, **, ***: significant at 5%, 1%, and 0.1%

Table A3. Air Pollution and Traffic Accidents Caused by Violations (IV and Control Function)

| Variables | (1) IV PPML Coefficient | SE | (2) IV PPML Coefficient | SE | (3) IV PPML Coefficient | SE | (4) IV OLS Coefficient | SE |
|---|---|---|---|---|---|---|---|---|
| $PM_{2.5}$ | -0.0013 | (0.0065) | -0.0056*** | (0.0012) | -0.0059*** | (0.0012) | -0.0070*** | (0.0016) |
| Temp. (°C) | | | 0.0071 | (0.0058) | 0.0013 | (0.0055) | -0.0020 | (0.0074) |
| [Temp. (°C)]² | | | -0.0000 | (0.0001) | 0.0001 | (0.0001) | 0.0002 | (0.0002) |
| Rainfall (mm/hr) | | | -0.0230** | (0.0071) | -0.0211** | (0.0070) | -0.0276*** | (0.0078) |
| [Rainfall (mm/hr)]² | | | -0.0013 | (0.0009) | -0.0014 | (0.0009) | -0.0005 | (0.0006) |
| Rain hours (hr) | | | -0.0044** | (0.0014) | -0.0045** | (0.0015) | -0.0056** | (0.0017) |
| [Rain hours (hr)]² | | | -0.0000 | (0.0001) | -0.0000 | (0.0001) | -0.0001 | (0.0001) |
| Relative Humidity (%) | | | 0.0036 | (0.0038) | 0.0037 | (0.0040) | -0.0003 | (0.0054) |
| [Relative Humidity (%)]² | | | -0.0001* | (0.0000) | -0.0001* | (0.0000) | -0.0000 | (0.0000) |
| Wind Speed (m/s) | | | 0.0263* | (0.0106) | 0.0071 | (0.0098) | 0.0068 | (0.0115) |
| [Wind Speed (m/s)]² | | | -0.0044** | (0.0014) | -0.0022 | (0.0013) | -0.0016 | (0.0010) |
| Constant | 0.2561 | (0.1772) | 1.0812*** | (0.1526) | 1.1928*** | (0.1497) | 1.6914*** | (0.2168) |
| | | | | | | | | |
| Observations | 892,044 | | 886,182 | | 820,358 | | 888,736 | |
| Weather Controls | N | | Y | | Y | | Y | |
| Town FE | N | | Y | | N | | N | |
| Year-Month FE | N | | Y | | N | | N | |
| DoW FE | N | | Y | | N | | N | |
| Town-Year-Month FE | N | | N | | Y | | Y | |
| Town-DoW FE | N | | N | | Y | | Y | |
| K-P F-statistics | 99.67 | | 46.27 | | 31.31 | | 31.31 | |

Notes: This table presents complete results of table 4. The outcome variable is the total number of accidents caused by violations in each district/township. Standard errors are clustered at number of day and district/township level.

*, **, ***: significant at 5%, 1%, and 0.1%

Table A4. Air Pollution and Traffic Volume (IV and Control Function)

| Variables | (1) Linear Coefficient | SE | (2) Quadratic Coefficient | SE | (3) Splines Coefficient | SE |
|---|---|---|---|---|---|---|
| $PM_{2.5}$ | 0.0047 | (0.0035) | 0.0047 | (0.0033) | | |
| $(PM_{2.5})^2$ | | | 0.0000 | (0.0000) | | |
| $PM_{2.5}$ (0 – 10 µg/m³) | | | | | 0.0029 | (0.0036) |
| $PM_{2.5}$ (10 – 20 µg/m³) | | | | | 0.0062 | (0.0034) |
| $PM_{2.5}$ (20 – 30 µg/m³) | | | | | 0.0021 | (0.0050) |
| $PM_{2.5}$ (30 µg/m³ and above) | | | | | 0.0072 | (0.0034) |
| Observations | 11,154 | | 11,154 | | 11,154 | |

Notes: The outcome variable is the logged daily total traffic volume by town (district) in Taipei City, 2018 to 2020. All models control for Town-Year-Month and Town-DoW fixed effects as well as weather controls including the linear and squared terms of daily average temperature, relative humidity, per hour rainfall, hours of rain, and wind speed. $PM_{2.5}$ concentration is instrumented by wind directions. The first-stage K-P F-statistic is 22.12. Standard errors are in parenthesis and clustered at number of day and district/township level. All the coefficients of $PM_{2.5}$ are not significant at 5% level.

Table A5. Air Pollution and Traffic Accidents Caused by Violations (IV and Control Function)

| Variables | (1) Linear Weather | | (2) Eight Wind Directions | | (3) Four Wind Direction and Wind Speed | | (4) County Level | | (5) Sunny and Good Visibility Only | |
|---|---|---|---|---|---|---|---|---|---|---|
| | Coefficient | SE | Coefficient | SE | Coefficient | SE | Coefficient | SE | Coefficient | SE |
| $PM_{2.5}$ | -0.0069*** | (0.0011) | -0.0055*** | (0.0009) | -0.0046*** | (0.0006) | -0.0069* | (0.0029) | -0.0061*** | (0.0018) |
| Observations | 820,358 | | 820,358 | | 820,358 | | 48,518 | | 804,604 | |
| Weather Controls (Linear terms) | Y | | Y | | Y | | Y | | Y | |
| Weather Controls (Quadratic terms) | N | | Y | | Y | | Y | | Y | |
| Town-Year-Month FE | Y | | Y | | Y | | N | | Y | |
| Town-DoW FE | Y | | Y | | Y | | N | | Y | |
| County-Year-Month FE | N | | N | | N | | Y | | N | |
| County-DoW FE | N | | N | | N | | Y | | N | |
| K-P F-statistics | 34.02 | | 19.27 | | 55.39 | | 23.95 | | 31.31 | |

Notes: The outcome variable is the total number of accidents caused by violations in each district/township. Model 1 includes only linear weather controls of daily average temperature, relative humidity, per hour rainfall, hours of rain, and wind speed as well as Town-Year-Month and Town-DoW fixed effects. Model 2 uses instrumental variables with eight wind directions and includes both linear and squared terms of weather controls. Model 3 uses wind speed and four wind directions as the excluded instruments. Model 4 uses daily observations collapsed into county level. Model 5 uses the daily number of accidents caused by violations and occurred in sunny days with good visibility at each district/township as the outcome variable. Standard errors in model 1, 2, 3, and 5, are clustered at number of day and district/township level, while those in model 4 are clustered at number of day and county/level.

*, **, ***: significant at 5%, 1%, and 0.1%

Table A6. Ozone and Traffic Accidents Caused by Violations (IV and Control Function)

| Variables | (1) Ozone Coefficient | SE |
|---|---|---|
| Ozone (ppm) | 0.0003 | (0.0008) |
| | | |
| Observations | 820,345 | |
| Weather Controls | Y | |
| Town-Year-Month FE | Y | |
| Town-DoW FE | Y | |
| K-P F-statistics | 57.62 | |

Notes: The outcome variable is the total number of accidents caused by violations in each district/township. The air pollutant variable is daily average of ozone concentration (ppm) estimated by inverse distance weighting of the readings from three nearest air quality monitoring stations of each district/township. Weather controls include the linear and squared terms of daily average temperature, relative humidity, per hour rainfall, hours of rain, and wind speed. Standard errors are in parenthesis and clustered at number of day and district/township level.